\documentclass[pra,onecolumn,amsmath,amssymb]{revtex4-2}
\usepackage{bm}% bold math
\usepackage{epsfig}
\usepackage{mathrsfs}

\begin{document}

\title{Variational Principles for Hierarchical Systems of Equations} 

\date{\today}
 
\author{A.\  R.\  P.\  Rau\footnote{ravirau@lsu.edu} \\
Department of Physics and Astronomy, 
Louisiana State University, 
Baton Rouge, Louisiana 70803}

\begin{abstract}
A unified construction of variational principles (VP) is applied to hierarchical systems of equations. The adjoint set of equations involved in this construction differ from the original hierarchy set in important ways. Notably, upward and downward nature of hierarchy are reversed between the two sets, as are boundary conditions and, most significantly, whether or not the equations are factorizable. These features may make for ease of solving one of the two sets and thereby for usefulness of the VP themselves that employ both. 
 
\end{abstract}

%\pacs{03.65.Ge, 03.65.Sq, 31.15.xj}

\maketitle
    
%\begin{specialtable}[H]

\section{Introduction}
Variational Principles \cite{ref1,ref2} are among the most powerful of approximation techniques in mathematical physics and engineering. When available, they are often more useful and easier to use than other approximation techniques such as perturbation series. First courses in classical and quantum physics teach this. In many particle quantum physics, the Rayleigh-Ritz variational method for the ground state energy of a complex system is often the only technique available and, as a result, ubiquitously used. This principle is based on ``the physics of the ground state," that it is the lowest energy state. Thus, of all possible guessed or ``trial" functions for this state, the one that yields the lowest energy is the best of such a class of states so that, introducing trial parameters and varying them to reach that lowest energy is a way of converging closer to the exact ground state. In this case, not only is the energy expression ``variational" or ``stationary" (that is, first derivatives vanish) but also its difference from the exact value of well defined sign so that it is an upper bound on the ground state energy, an ``extremum." Other VPs are often stationary and not extremum but, nevertheless, also useful in that the ``errors" are of second order or higher when compared with the first-order error in the guess from the exact solution. Thus, once ``close" to the exact value, successive steps become more accurate. 

Apart from the Rayleigh-Ritz, other VPs are rare with a few well-known ones in scattering theory attributed to the genius of the discoverers whose names are attached as with Rayleigh and Ritz: Schwinger, Kohn, etc. A unified construction \cite{ref2,ref3,ref4} of VPs for very general systems of equations showed, however, that the basic philosophy of Lagrange undetermined multipliers which is well known in the Euler-Lagrange-Hamilton formulation provides a clue to how to construct VPs for any given set of equations whatever their nature, linear or nonlinear, differential or integral, etc. The basic idea is to start with the defining set of equations for the unknown functions involved in the desired functional of those functions, write that desired functional with guessed solutions and add to it the defining set with Lagrange functions (more general than constants as in usual Euler-Lagrange variational functionals) as ``constraints." With guessed solutions for both the original set and the adjoint set of exact Lagrange functions (also initially unknown), that resulting functional can be ``made," ``engineered," to have no first order errors. This step defines the adjoint set and gives an expression guaranteed by construction to have no first-order errors in either original or adjoint set. Thus a variational/stationary estimate is obtained for what is being sought. Many examples have been given of this procedure \cite{ref2}, including application to nonlinear equations \cite{ref5} when the adjoint equations are by construction linear in the Lagrange functions and thus perhaps more amenable to guessed solutions.

However, one class of equations not yet subject to this general construction are systems of hierarchical equations that occur especially in statistical physics, turbulence, and other such problems involving connections across different steps of a hierarchy, whether in scales or many-particle density or other correlation functions \cite{ref6,ref7,ref8,ref9,ref10}. This is the subject of this paper, to examine the nature of the differences between original and adjoint sets and exploit simplicities that may be in one but not the other. Indeed, we find some, notably in reversing upward hierarchy of the original with resulting integral couplings involved, to downward and thus factorizable structures in the adjoint. Alongside, boundary conditions at the origin are exchanged for those at infinity.

\section{A simple hierarchy of differential equations}
 Consider as a simple, opening example, the set of second-order differential equations
 
 \begin{eqnarray}
 (y_1(x))'' & = & y_2(x), \\ \nonumber
 (y_2(x))'' & = & y_3(x), \\ \nonumber
 & \ldots \,\, ,&
 \label{eqn1}
 \end{eqnarray}
 with boundary conditions $y_i(0) =0, y_i^{'}(0) =1$ where primes represent differentiation with respect to $x$. In solving for $y_i(x)$ in such an upward hierarchy of a large set, $i=1, 2, \ldots N$, one may be tempted to truncate the hierarchy by dropping higher $y_i$ but that would be inconsistent even with the boundary conditions.
 
Instead, the general construction \cite{ref2} of a VP for the set ${y_i(x)}$ would use a ``trial" or guessed set of known ${y_{it}(x)}$ satisfying the boundary conditions, with the subscript $t$ denoting trial, to write
 
\begin{eqnarray}
\left( 
\begin{array}{c}
y_1(x) \\  
y_2(x)\\  
. \\.  
. \\.  
.
\end{array}
\right)_v
 & = &
\left( 
\begin{array}{c}
y_1(x) \\ 
y_2(x)\\ 
. \\
. \\
.
\end{array}
\right)_t
+
\int_0^x ds
\left(
\begin{array}{ccccc}
L_{11}(x,s) & L_{12}(x,s) & . & . & . \\
L_{21}(x,s) & L_{22}(x,s) & . & . & . \\
. & . & . & . & . \\
. & . & . & . &.  \\
. & . & . & . & .
\end{array}
\right)_t 
\times \\ \nonumber
& & 
\left[
\left(
\begin{array}{ccccc}
d^2/ds^2 & 0 & 0 & 0 & 0 \\
0 & d^2/ds^2 & 0 & 0 & 0 \\
0 & 0 & d^2/ds^2 & 0 & 0  \\
0 & 0 & 0 & d^2/ds^2 & 0  \\
0 & 0 & 0 & 0 & d^2/ds^2 
\end{array}
\right)
-
\left(
\begin{array}{ccccc}
0 & V_1(s) & 0 & 0 & 0 \\
0 & 0 & V_2(s) & 0 & 0 \\
0 & 0 & 0 & V_3(s) & 0  \\
0 & 0 & 0 & 0 & V_4(s)  \\
0 & 0 & 0 & 0 & 0
\end{array}
\right)
\right]
\left( 
\begin{array}{c}
y_1(s) \\ 
y_2(s)\\ 
. \\
. \\
.
\end{array}
\right)_t
\label{eqn2}
\end{eqnarray}

The expression in Eq.(2) satisfies the requirement that the right-hand side is computable, involving only known trial functions. It also clearly gives the exact solution if the $(y_i(x))_t$ happen to be the exact solutions regardless of what the generalized Lagrange multipliers $L_{ij}(x,s)_t$ are. They are the trial solutions of an adjoint set of equations for adjoint functions $L$ that are at this point unknown but are now defined by setting all first-order errors on the right-hand side equal to zero, with the ``definition" of first-order errors as $y_{it} = y_i + \delta y_i, \,\, L_{ijt} = L_{ij} + \delta L_{ij}$. Formally, this amounts to setting 

\begin{equation}
\delta \mathscr{Y} + \int_0^x ds \,\, \mathscr{L}(x,s) [\frac{d^2}{ds^2} \mathscr{I} - \mathscr{V}(s)]\, \delta \mathscr{Y} =0.
\label{eqn3}
\end{equation}

Transferring the two differentials with respect to $s$ through integration by parts and using the fact that $\delta \mathscr{Y} (0) =0$ gives the differential equation satisfied by the Lagrange functions

\begin{equation}
\frac{d^2}{ds^2} \mathscr{L} (x,s) -\mathscr{L} (x,s) \mathscr{V}(s) =0,
\label{eqn4}
\end{equation}
together with the boundary conditions 

\begin{equation}
\mathscr{L} (x,x) =0, \,\,\,\, [d\mathscr{L} (x,s)/ds ]_{s=x} =\mathscr{I},
\label{eqn5}
\end{equation}
that is, $L_{ij}(x,x)=0, \,\, L_{ij}'(x,x) =\delta_{ij}$. Note that primes here are with respect to the second index $s$. 
As elsewhere in such a VP construction \cite{ref2,ref5}, adjointness reverses the nature of equations and boundary conditions. Whereas the original defining equations for $y_i$ that may now be depicted as 

\begin{equation}
\frac{d^2}{ds^2} \mathscr{Y} = \mathscr{V}(x) \mathscr{Y}(x), \,\,\,\, \mathscr{Y} (0)=0, \mathscr{Y}'(0)=1,
\label{eqn6}
\end{equation} 
satisfy boundary conditions at $x=0$, the adjoint set $\mathscr{L}(x,s)$ have boundary conditions at the ``opposite" end of $s=x$, and while the original set in Eq.(6) are an upward hierarchy, the adjoint set in Eq.(4) is downward, that is,

\begin{equation}
L_{11}''=0, \,\,L_{12}''=V_1L_{11},\,\, L_{13}''=V_2L_{12},\, \ldots \,L_{i1}''=0,\,\, L_{ij}''=V_{j-1}L_{i,j-1}.
\label{eqn7}
\end{equation}

In this admittedly simple example, the adjoint set may even be solved exactly:

\begin{equation}
L_{11} = s-x, \,\, L_{12}''(s) =V_1(s) (s-x), \ldots, L_{21}=0,\,\, L_{22} =s-x,\,\, L_{23}'' =V_3(s) (s-x), \ldots \,\, .
\label{eqn8}
\end{equation}
No trial or guessed solution of these $\mathscr{L}$ is needed and when substituted into the VP in Eq.(2), linearity will ensure an exact value for any trial $(\mathscr{Y})_t$. We have what may be called a variational identity \cite{ref4}. More generally, however, the adjoint set may not admit exact solution but, nevertheless, its downward nature may give more accessible trial solutions and then the VP will provide the advantage that the variational estimate on its left-hand side has no first order $\delta\mathscr{Y}$ but only second-oder errors.  

Rate equations such as in radioactive decay chains of mother-daughter generations are examples of such linear hierarchical equations and the VP approach can be fruitfully applied to them. It is a simple exercise to show that in a case with constant decay and regeneration coefficients for such a decay chain, even the simplest trial function satisfying only the required boundary/initial condition of starting with parent population unity, all others zero, will generate the exact solution of exponential decay of that initial parent and more complicated combinations of exponential dependence on time of subsequent members in the decay chain.

As another example, consider the linear set of equations such as ones that occur in deep x-ray penetration in matter \cite{ref11}

\begin{equation}
\sum_{m=1}^n a_{nm}x_m +px_{n+1} =c_n, \,\, n=1, 2, \ldots N, \,\, x_{N+1}=0,
\label{eqn9}
\end{equation}
with $(a,p,c)$ coefficients, possibly constants. Again, such an upward hierarchy will require for solving in sequence an initial assumption on $x_1/x_2$ and proceeding to the end $x_{N+1}$ on which the boundary condition has to be applied, iterating with different input to satisfy that final desired value. Once again, the downward hierarchy of $L$ yields to an easy closed form solution.

\section{Hierarchical Green's functions}
Next, consider the set of equations more complicated than Eq.(1)

\begin{eqnarray}
\frac{d}{dx}y_1(x) & = & \int dx' \,\,V_1(x') y_2(x, x'), \\ \nonumber
(\frac{d}{dx}+\frac{d}{dx'})y_2(x, x') & = & \int dx'' \,\, V_2(x'') y_3(x, x', x''), \\ \nonumber
 & \ldots \,\, ,&
 \label{eqn10}
 \end{eqnarray}
typical of many hierarchical structures in physics \cite{ref6,ref7,ref8}. The primes on $x$ are not derivatives but stand for alternative values of that one-dimensional coordinate. The $y$ may be different many-particle densities, the one-particle density dependent on a two-particle density with necessarily an integration involved, etc. Leaving aside for a moment the initial conditions on $y$ needed for such first-order differential equations, again in seeking a VP, we would write 
 
\begin{eqnarray}
y_{1v} (x) & = & y_{1t} (x)+ \int dx' L_{1t} (x, x') [ dy_{1t}(x')/dx' -\int dx'' V_1(x'')\,\,y_{2t}(x', x'')] + \int dx' \int dx'' L_{2t} (x, x', x'') \times \\ \nonumber
& & [dy_{2t}(x', x'')/dx' + dy_{2t}(x', x'')/dx''-\int dx''' V_2(x''')\,\,y_{3t}(x', x'', x''')] + \ldots .
\label{eqn11}
\end{eqnarray}
If all integrals are from 0 to $x$, and initial conditions given are $y_1(0), y_2(x,0), y_2(0,x)$, etc., then the one integration by parts of each term to derive the adjoint set of equations satisfied by the $L$, lead to boundary conditions on them at the opposite value of final $x$,

\begin{equation}
L_1(x,x) =-1, \,\, L_2(x, x'=x, x'') =\,L_2(x, x', x''=x) = 0, \ldots ,
\label{eqn12}
\end{equation}
and to the adjoint set of equations
 
\begin{equation}
dL_1(x,x')/dx' =0 \,\, or \,\,  \delta(x-x'), \,\,\, (d/dx +d/dx') L_2(x, x', x'') =-V_1(x'') L_1(x, x'), \, \ldots .
\label{eqn13}
\end{equation}
Note that because of the first condition in Eq.(12), the function $L_1(x, x')$ has a jump discontinuity at $x=x'$. And, successive equations for $L_i$ being a downward hierarchy necessarily have factorizable right-hand sides unlike the integral forms in the original set in Eq.(10). This is a necessary consequence of the adjointness inherent to such general VPs. Once again, they mean that the adjoint set in Eqs.(12,13) is easier to solve, or at least get trial solutions of for insertion into the VP in Eq.(11), than might be possible of the original set for $y$. 
 
\section{The BBGKY Hierarchy}
Starting from its discussion in statistical physics and fluid mechanics, the BBGKY (Bogoliubov-Born-Green-Kirkwood-Yvon) equations for $N$-particle distribution functions have seen widespread application in physics \cite{ref6,ref7,ref8,ref9,ref10}, including to galaxy clusters \cite{ref12,ref13} and the QCD gas \cite{ref14}. Using for simplicity one-dimensional coordinates and velocities, although immediate generalizations to vectors, gradients, and dot products are straightforward, an $N$-particle distribution function $f_N(r_1, \ldots r_N; v_1, \ldots v_N; t)$ and its family of reduced distribution functions \cite{ref8}

\begin{equation}
f_n(r_1, \ldots r_n; v_1, \ldots v_n; t) = \frac{N!}{(N-n)!} \int \Pi_{i=n+1}^N dr_i \,\,dv_i \,\,f_N(r_1, \ldots r_N; v_1, \ldots v_N; t),
\label{eqn14}
\end{equation} 
satisfy the BBGKY tower of hierarchical equations (see, for instance, Sec. 7.4 of \cite{ref7} or Eqs.(4.28, 29) of \cite{ref8})

\begin{eqnarray}
\frac{\partial f_1(r_1, v_1)}{\partial t} + v_1 \frac{\partial f_1}{\partial r_1} & = & \int \int dr_2 \,dv_2 \,\, \frac{\partial V_{1,2}}{\partial r_1} \frac{\partial f_2(r_1, r_2; v_1, v_2)}{\partial v_1}, \\ \nonumber
\frac{\partial f_2(r_1, r_2; v_1, v_2 )}{\partial t} + v_1 \frac{\partial f_2}{\partial r_1} + v_2 \frac{\partial f_2}{\partial r_2} & = & \int \int dr_3 dv_3 \,\,\left( \frac{\partial V_{1,3}}{\partial r_1} \,\,\frac{\partial}{\partial v_1} + \frac{\partial V_{1,3}}{\partial r_2} \,\,\frac{\partial}{\partial v_2}\right)  f_3(r_1, r_2, r_3; v_1, v_2, v_3), \\ \nonumber
 & \ldots , & \\ \nonumber
 & \ldots & \,\,\,\,\,\, .
\label{eqn15}
\end{eqnarray}

Practical use of such a complex open hierarchy necessarily truncates (Sec. 4.4 of \cite{ref8}) at some $f_n$ and even then has to assume some kind of Boltzmann-like ansatz \cite{ref7,ref8} such as writing $f_2 (r_1, r_2; v_1, v_2)= f_1(r_1, v_1) f_1(r_2, v_2)$ to close the set at the very first step or, perhaps, write $f_2 = f_1 f_1 +g_{12}$ with a ``correlation" function $g_{12}$ assumed to be small and setting the right-hand side of the next equation to zero (Eq.(4.37) of \cite{ref8}). Such admittedly {\it ad hoc} approximation schemes are legion. See an example of the latter in a recent treatment of a weakly confined QCD gas \cite{ref13}. Applications for galaxy clusters have argued for a useful closing of the hierarchy by assuming that higher point correlations are appropriately symmetrized sums of products of two-point correlations. See the so-called Kirkwood superposition in Eq.(7.4.8) of \cite{ref7} that decomposes $f_3$ as a triple product of $f_2$ leading to the nonlinear Born-Green-Yvon Eq.(7.4.8). Another alternative is in Eq.(4.38) of \cite{ref8}. These approximations also have the attendant effect of rendering the equations nonlinear unlike the original complete hierarchy which is linear in the $n$-particle distribution functions, albeit hierarchically coupled upward. 

Instead, consider treating the hierarchy in Eq.(15) to a VP as in previous sections. In order to use $t$ as a subscript for trial solutions, we will remove explicit reference to time from now on, replacing its occurrence above by the frequency parameter through $i\omega$, and write

\begin{eqnarray}
f_{1v} (r_1, v_1) & = & f_{1t} (r_1, v_1) +\int \! dr_1' \int \! dv_1' L_{1t} (r_1, v_1, r_1', v_1') \left( i\omega f_{1t} (r_1', v_1') + v_1' \frac{\partial f_{1t}}{\partial r_1'} -  \int  dr_2' \int dv_2' \frac{\partial V_{1', 2'}}{\partial r_1'} \frac{\partial f_{2t}(r_1', r_2', v_1', v_2')}{\partial v_1'} \right) \\ \nonumber
& & + \! \int \! dr_1' \!\int \! dr_2' \int \! dv_1' \int \! dv_2' \, L_{2t} (r_1, r_1', r_2', v_1, v_1', v_2') [ i \omega f_{2t}(r_1', r_2', v_1', v_2') + \!v_1' \frac{\partial f_{2t} (r_1', r_2', v_1', v_2')}{\partial r_1'} \!+ v_2' \frac{\partial f_{2t} (r_1', r_2', v_1', v_2')}{\partial r_2'}  \\ \nonumber
& & -\int dr_3' \! \int dv_3' \left( \frac{\partial V_{1',3'} }{\partial r_1'} \frac{\partial}{\partial v_1'} + \frac{\partial V_{2',3'} }{\partial r_2'} \frac{\partial}{\partial v_2'} \right) f_{3t} (r_1', r_2', r_3', v_1', v_2', v_3') ]  \\  \nonumber
& & + \ldots \,\,\, .
\label{eqn16}
\end{eqnarray}

As usual, next writing trial quantities as $f_{1t} = f_1 +\delta f_1$, etc., transferring through integration by parts derivatives on $f$ to derivatives on $L$, and setting all coefficients of $\delta f$ to zero gives the equations defining the adjoint Lagrange functions:

\begin{eqnarray}
i\omega L_1(r_1, v_1, r_1', v_1') -(\partial/\partial r_1') v_1' L_1 (r_1, v_1, r_1', v_1') & = & - \delta(r_1-r_1') \delta (v_1-v_1'), \\ \nonumber
i\omega L_2(r_1,v_1,r_1',v_1',r_2',v_2') -(\partial/\partial r_1')v_1'L_2 -(\partial/\partial r_2') v_2'L_2 & =& -\frac{\partial}{\partial v_1'} \,\,\frac{\partial V_{1',2'}}{\partial r_1'} \,L_1(r_1,v_1,r_1',v_1'), \\ \nonumber
& & \ldots \,\,\, .
\label{eqn17}
\end{eqnarray}
The boundary conditions are that the $L$ functions vanish at those end points of primed variables where the corresponding $f$ are not fixed so that values of $\delta f$ there are unknown. Note the reversed nature of the hierarchy in $L$ equations with the first, $L_1$  essentially a Green's function with only Dirac delta terms on the right-hand side, the $L_2$ equation on the other hand having no $v_2'$ dependence at all on its right-hand side, etc. The next of the equations continues this pattern with no $v_3'$ dependence on its right-hand side

\begin{equation}
i\omega L_3(r_1,v_1,r_1',v_1',r_2',v_2',r_3',v_3')-[(\partial/\partial r_1') v_1'+(\partial/\partial r_2') v_2'+(\partial/\partial r_3') v_3']L_3=-[\frac{\partial}{\partial v_1'} \frac{\partial V_{1',3'}}{\partial r_1'}+\frac{\partial}{\partial v_2'} \frac{\partial V_{2',3'}}{\partial r_2'}]L_2.
\label{eqn18}
\end{equation}

It is interesting to recognize the adjoint functions and their equations in Eq.(17,18). The first of them for $L_1$ is the same as the Green's function $L_1= \mathscr{G}^{(0)} (rvt|r'v't')$ in Eq.(4.8) of \cite{ref6}.  Its immediate solution is, again as in Sec. 5.10, 11 of \cite{ref6},

\begin{equation}
L_1(r_1, v_1. r_1', v_1') = \delta (v_1 -v_1') e^{i\omega r_1'/v_1'} \left( A(r_1) + e^{-i\omega r_1/v_1}\right).
\label{eqn19}
\end{equation}
Next, the implication for $L_2$ , because the right-hand side of its defining equation has $(\partial/\partial v_1') \delta(v_1-v_1')$, that $L_2$ also is proportional to it. Thus, the sequence $L_i$ involve the Dirac delta function and its successive derivatives.

Fourier transforming the equations for $L_i$, 
\begin{eqnarray}
L_1 & = & \delta (v_1-v_1') \int \, dk_1 \,\,\tilde{L_1} (\omega, k_1) e^{ik_1(r_1-r_1')}, \\ \nonumber
L_2 & = & (\partial/\partial v_1') \delta(v_1-v_1') \int dk_1 \int dk_2 \,\, \tilde{L_2}(\omega, k_1, k_2, v_1, v_1', v_2') e^{ik_1(r_1-r_1')} e^{ik_2(r_1'-r_2')}, \\ \nonumber
& \ldots &,
\label{eqn20}
\end{eqnarray}
with

\begin{eqnarray}
\tilde{L_1}(\omega, k_1) & = & \frac{-1}{i(k_1 \cdot v_1 + \omega)}, \\ \nonumber
\tilde{L_2}(\omega, k_1, k_2, v_1, v_1' , v_2') & = & \frac{k_2 \cdot V(k_2)/(k_1 \cdot v_1 +\omega)}{i(\omega + (k_1-k_2) \cdot v_1' +k_2 \cdot v_2')}, \\ \nonumber
& \ldots \,\,\, .& 
\label{eqn21}
\end{eqnarray}

\section{Summary}
Variational or stationary principles for a quantity of interest in a problem that is not exactly solvable provide an attractive approximation scheme. Through successive iteration, often with open numerical parameters that are themselves varied, the feature that the variational expression contains only second-order errors serves to converge towards the desired solution. A unified view of VPs regards the original set of equations describing the system as constraints through undetermined Lagrange functions that are then tailored to satisfy equations defining an adjoint set to the original. Approximate or trial solutions of both sets then evaluate the quantity of interest through the VP. The adjoint set brings complementary aspects to the original which explains the possible power of a VP that combines both. For instance, in initial value problems that integrate out from some initial condition out to some distance, portions of the forces and potentials not yet seen influence through the adjoint set that has its boundary condition at the opposite end and carries in information inward from that point \cite{ref5}. Applied to hierarchical equations such as the ones of statistical physics we consider here, the upward nature of the original hierarchy is also reversed into a downward aspect in the adjoint Lagrange set. (See also forward-backward treatments \cite{ref15,ref16} in Kolmogorov equations that also involve such a change in sign.) This often has the accompanying feature that integro-differential terms in the original hierarchy manifest as factorized source terms in the adjoint. Again, this may afford an easier set of approximate solutions of that set and thereby an improvement in the VP result compared to just solving approximately the original set. Earlier work has shown that with nonlinear equations, differential or integral, the fact that the adjoint set is linear as a matter of construction also affords a simplification. Combined, this suggests that nonlinear hierarchies may also be fruitfully investigated in our VP framework.

\end{document}